\documentstyle[twocolumn,aps,prl,floats,epsf,multirow]{revtex}

\newcommand\beq{\begin{equation}}
\newcommand\eeq{\end{equation}}
\newcommand\beqn{\begin{eqnarray}}
\newcommand\eeqn{\end{eqnarray}}
\newcommand{\la}{\langle}
\newcommand{\ra}{\rangle}

\begin{document}

\draft

\wideabs{

\title{Nonperturbative Gluon Radiation 
and Energy Dependence of Elastic Scattering}
\author{B.Z.~Kopeliovich$^{1,2}$,
I.K. Potashnikova$^{1,2}$,
B.~Povh$^{1}$
and E.~Predazzi$^{3}$
\vspace*{9pt}}
\address{
$^{1}$ Max-Planck Institut
f\"ur
Kernphysik,
Postfach
103980, 69029 Heidelberg,
Germany\\
$^{2}$ Joint
Institute
for Nuclear Research, Dubna,
141980
Moscow Region,
Russia\\
$^3$ Universit\`a di
Torino
and INFN, Sezione di Torino, I-10125, Torino,
Italy}

\maketitle

\begin{abstract}
The energy dependence of the total hadronic cross sections is
caused by gluon bremsstrahlung which we treat nonperturbatively.
It is located at
small transverse distances about $0.3\,fm$ from the valence quarks.
The cross section of gluon radiation 
is predicted to exponentiate and rise with
energy as $s^{\Delta}$ with $\Delta=0.17\pm 0.01$.
The total cross section also includes a large energy independent
Born term which corresponds to no gluon radiation. The calculated
total cross section and the slope of elastic scattering
are in good agreement with the data.

\end{abstract}


} 


The dynamics of energy dependence of the hadronic total 
cross sections is a long standing challenge since 1973 when
this effect was first observed at the ISR. 
In DIS the source for the rising total cross section for interaction
of highly virtual photons is well understood in QCD as
caused by an intensive gluon bremsstrahlung \cite{book,bfkl}.
Indeed, radiation of each gluon supply an extra 
${\rm ln}s$ and ${\rm ln}Q^2$. This is a specific regime
of radiation when a $\bar qq$
fluctuation of the photon of a tiny size $\sim 1/Q$ 
radiates gluons at much larger 
transverse separations. 

It is difficult to extend the perturbative results to  
soft hadronic collisions because it is
quite a different regime where the approximations made
in the perturbative case break down. 
Namely, gluon radiation giving rise to the energy dependence
of the total cross section occurs at rather small transverse
distances around the valence quarks, $r_0\approx 0.3\,fm$
which are much smaller than the mean interquark spacing in light 
hadrons. This conclusion follows from the analysis \cite{kst2} 
of the data for diffractive gluon radiation based on
the light-cone approach when the effective nonperturbative
interaction of radiated gluons is included. 

The smallness of the gluon clouds of the valence quarks 
is confirmed by the study of the
gluon formfactor of the proton employing 
QCD sum rules \cite{braun}. 
The $Q^2$ dependence of
the formfactor turns out to be rather weak 
corresponding to a small radius of the gluon
distribution which was estimated
at the same value $r_0\approx 0.3\,fm$.

Another evidence for a short gluon-gluon correlation length
$\lambda \approx 0.3\,fm$ arises in
the stochastic vacuum model of Dosch and Simonov 
\cite{dosch,ds}, as it was measured on the
lattice \cite{pisa}. 
In the case of a poorly populated gluon cloud (only about one gluon
is radiated by a valence quark at available energies, see below)
this corresponds to 
the correlation radius between the quark and the gluon.

The same size $\sim 0.3\,fm$ 
emerges from the Shuryak's instanton liquid model \cite{shuryak}
as the instanton size which controls the mean radius of the sea 
surrounding a valence quark and by many phenomenological analyses.

In the Gribov's theory of confinement \cite{gribov,ewerz}
the same distance $\sim 0.3\,fm$ should correspond to
the critical regime related to breaking of chiral symmetry.
Namely, at smaller distances, a perturbative quark-gluon basis
is appropriate, while at larger separations quasi-Goldstone
pions emerge. The corresponding critical 
value of the QCD running constant $\alpha_c=0.43$
evaluated in \cite{gribov} turns out
to be very close to our estimate (see below)
of $\alpha_s$ corresponding
to gluon radiation separated by $0.3\,fm$.
The value of $\alpha_s$ is crucial
for our evaluation of the energy dependence
of the gluon bremsstrahlung.

It is quite plausible that all these
observations are the manifestations of
the same dynamics, however it is still unclear
how to make a Lorentz boost in these approaches.
This is the advantage of the light cone treatment of
nonperturbative gluon radiation \cite{kst2} which seems to be
best designed for calculating the energy dependence of
the total cross section. We believe that the 
nonperturbative interaction of gluons introduced 
in \cite{kst2} as a light-cone potential is an effective
manifestation of properties of the QCD vacuum. 
Similar scale $\sim 0.3\,fm$ found in all these approaches
supports this conjecture.

An interesting attempt to implement the nonperturbative
gluon interaction into the Pomeron ladder building was
made recently by Kharzeev and Levin \cite{kl}
and Shuryak \cite{sh}. They found that the radiation of 
colorless pairs of gluons is a part of the leading-log 
approximation since each extra power of the coupling 
$\alpha_s$ cancels due to the strong glue-glue interaction.
The radiated glueballs are not clustering around the valence
quarks, but spreading all over the hadron.  
The estimated $\Delta \approx 0.05$ \cite{sh} is about twice
as small (and even more so if corrected for unitarity) as the data need.
Although the scale for
$\alpha_P^{\prime}\sim 1/M_0^2$ seems to be correct,
an extra factor $\Delta/4$ makes it too small.

We start calculating the energy dependence of the  
total cross section summing up the
contributions of different Fock components of the incident hadron,
\beq
\sigma^{hN}_{tot}=\sum\limits_{n=0}\sigma^{hN}_n\ .
\label{6.3a}
\eeq
To avoid double-counting, we sum over cross sections $\sigma_n$ 
of physical processes corresponding to the radiation of $n$ gluons.

The lowest Fock component of a hadron 
contains only valence quarks. The corresponding Born term in the total
cross
section has the form (for the sake of simplicity we assume that 
the incident hadron is a meson),
\beq
\sigma^{hN}_0 = \int\limits_0^1 d\alpha_q
\int d^2 R\,\left|\Psi^h_{\bar qq}(\alpha_q,R)\right|^2\,
\sigma^N_{\bar qq}(R)\ .
\label{6.4}
\eeq
Here the Fock state wave function 
$\Psi^h_{\bar qq}(\alpha_q,R)$ depends 
on the transverse $q-\bar q$ separation $R$ and on
the fraction $\alpha_q$ of the light-cone momentum of the pair
carried by the quark.
The cross section $\sigma^N_{\bar qq}(R)$ of interaction of
the valence $\bar qq$ dipole with a nucleon 
cannot be calculated perturbatively since the separation $R$ is
large. According to \cite{pirner} this energy independent term
has no relation to the smallness of the spots (gluon clouds)
in the hadron.

The next contribution to $\sigma_{tot}^{hN}$ comes from the
radiation of a single gluon. The radiation is possible only due to the
difference between the cross sections for the $\bar qq$ and $\bar qqG$ Fock
components, otherwise no new state can 
be produced \cite{kst2}. The cross section 
of radiation of a single gluon reads \cite{kst2},
\beqn
&&\sigma^{hN}_1 =
\int\limits_0^1 d\alpha_q \int d^2R\,\,
\Bigl|\Psi^h_{\bar qq}
(R,\alpha_q)\Bigr|^2
\nonumber\\ & \times & \,
{9\over4}\int\limits_{\alpha_G\ll 1}
\frac{d\alpha_G}{\alpha_G} 
\int d^2r
\biggl\{\Bigl|\Psi_{\bar qG}(\vec R 
+\vec r,\alpha_G)\Bigr|^2
\sigma_{\bar qq}^N(\vec R +\vec r) 
\nonumber\\ & + & \, 
\Bigl|\Psi_{qG}(\vec r,\alpha_G)\Bigr|^2
\sigma_{\bar qq}^N(r) -  
{\rm Re}\,\Psi_{qG}^*(\vec r,\alpha_G)\,
\Psi_{\bar qG}(\vec R +\vec r,\alpha_G)\,
\nonumber\\ & \times & \,
\Bigl[\sigma_{\bar qq}^N(\vec R +\vec r)+\sigma_{\bar qq}^N(r)-
\sigma_{\bar qq}^N(R)\Bigr]
\biggr\}
\label{6.5}
\eeqn
Here $\alpha_G$ is the fraction of the quark momentum carried by
the gluon, and $\vec r$ is the quark-gluon transverse separation.
The three terms in the curly brackets correspond
to the radiation of the gluon by the quark, by the antiquark 
and to their interference respectively.

The nonperturbative wave function for a quark-gluon Fock
component is derived in \cite{kst2}. Neglecting the
quark mass, the wave function reads,
\beq
\Psi_{qG}(\vec r,\alpha_G\ll 1)=
-\,\frac{2\,i}{\pi}\,\sqrt{\alpha_s\over3}\,\,
\frac{\vec e\,^*\cdot\vec r}{r^2}\,
e^{-r^2b_0^2/2}\ ,
\label{6.6}
\eeq
where $\vec e$ is the polarization vector of the massless gluon.
The parameter $b_0=0.65\,GeV$ characterizing the nonperturbative
quark gluon interaction is fixed by the data on large mass diffractive
dissociation corresponding to the triple-Pomeron limit.
It leads to quite a short mean quark-gluon separation
$r_0=\sqrt{\la r^2\ra}=1/b_0\approx 0.3\,fm$, which is small 
relative to the hadronic size. 
Therefore, only one or the other of the first two terms
in (\ref{6.5}) can be large, while the interference one can always be 
neglected. In this case, the integration in (\ref{6.5}) is easily
performed,
\beq
\sigma^{hN}_1 = N\,\frac{4\,\alpha_s}{3\,\pi}\,
\ln\left({s\over s_0}\right)\,\,
\frac{9\,C}{4\,b_0^2}\ .
\label{6.7}
\eeq
Here we assume that the approximation $\sigma_{\bar qq}^N(r)=
Cr^2$ is valid for $r\sim 1/b_0$.
$N$ is the number of valence quarks,
$\ln(s/s_0)=\ln[(\alpha_G)_{max}/(\alpha_G)_{min}]$,
where $(\alpha_G)_{min}=2\,b_0^2/s$, but  $(\alpha_G)_{max}$
is ill defined. It should be sufficiently small to 
use the wave functions (\ref{6.5}).
This leads to the condition to $s_0\gg 3\,GeV^2$. At high energy
$\sigma_1$ has little sensitivity on $s_0$ which
we fix at $s_0=30\,GeV^2$ for further applications.

The radiation of each new, $n$-th gluon can be treated as
radiation by a color triplet which is 
an effective quark surrounded by $n-1$ gluons.
It should be resolved by the soft interaction with the target
to be different from the radiation of $n-1$ gluons
{\it i.e.} the radiation cross section is proportional to the
difference between the total cross sections of the two subsequent
Fock states which is $9C/4b_0^2$. 
This can be also proved using a $1/N_c$
expansion and the dipole representation of Mueller \cite{al}. 
Since the radiation of a gluon with $\alpha_G \ll 1$ does not
affect the impact parameter of the radiating quark,
all the quark lines in the final state cancel with the 
same lines in the
initial state (see the prescription for calculating the radiative
cross section in \cite{kst1}), 
except for the radiation of the $n$-th gluon. 
Thus, $\sigma_n$ for quark-proton interaction 
in the leading-log approximation reads,
\beq
\sigma^{qN}_n = \frac{1}{n!}\,
\left[\frac{4\,\alpha_s}{3\,\pi}\,\,
{\rm ln}\left({s\over s_0}\right)\right]^n\,
\frac{9\,C}{4\,b_0^2}\ .
\label{6.7a}
\eeq

Summing up the powers of logarithms in (\ref{6.3a}) we arrive
at the following expression for the total cross section,
\beq
\sigma^{hp}_{tot}= \tilde\sigma^{hp}_0 + 
N\,\frac{9\,C}{4\,b_0^2}\,\,
\left({s\over s_0}\right)^{\Delta}\ ,
\label{6.8}
\eeq
with
\beq
\Delta=\frac{4\,\alpha_s}{3\,\pi}\ ,
\label{6.9}
\eeq
and $\tilde\sigma^{hp}_0=\sigma^{hp}_0-9C/4b_0^2$.
The soft Pomeron
intercept, $\alpha_P(0)=1+\Delta$, and can be evaluated
provided that the QCD coupling $\alpha_s$ is known. 

In Gribov's confinement scenario, chiral symmetry breaking
occurs when the running
coupling $\alpha_s$ exceeds the critical value
$\alpha_s=\alpha_c \approx 0.43$ \cite{gribov}.
This should happen at a distance of the order of
the size of a constituent quark $\sim 0.3\,fm$.
Therefore, this value can be used in (\ref{6.9}).

One can also calculate the mean $\la\alpha_s\ra$ 
for nonperturvative  gluon radiation averaging  over
transverse momenta $k_T$ of the radiated gluons.
The popular way to extend the running QCD coupling 
$\alpha_s(k_T^2)$ down to small
$k_T$ is a shift of the variable $k_T^2 \Rightarrow k_T^2
+ k_0^2$, where $k_0^2\approx 0.25\,GeV^2$
was evaluated in \cite{ewerz}
using the dispersive approach to calculating
higher twist effects in hard reactions \cite{dmw}.
The nonperturbative interaction of the radiated gluons
drastically suppresses small transverse
momenta, pushing $\la k^2_T\ra$ to higher values
which lowers $\alpha_s$.
We use the transverse momentum gluon distribution 
calculated in \cite{kst2}
in the light-cone approach in terms of
the universal color dipole cross section \cite{zkl}.
We calculated $\la\alpha_s\ra$ with a simple
parameterization 
$\sigma(\rho)\propto 1-{\rm exp}(\rho^2/\rho_0^2)$.
For a reasonable variation of $\rho_0= 0.3 - 1\,fm$
the mean coupling is in the range
$\la\alpha_s\ra = 0.38 - 0.43$ which is very close the 
the critical value mentioned above\cite{gribov}.
Taking the mid value $\la\alpha_s\ra= 0.4$ we get from
(\ref{6.9}),
\beq
\Delta = 0.17 \pm 0.01\ .
\label{6.11}
\eeq
This value is about twice as large as the
one suggested by the data for the energy dependence
of total hadronic cross
sections \cite{dl}. However, the
radiative part is a rather small fraction
of the total cross section (at medium high energies).
A structure similar to (\ref{6.8}) with a large $\Delta$
was suggested in \cite{knp} (with quite a different motivation)
and proved to agree well with the data.

The factor $C$
in the second term in (\ref{6.8}) can also be evaluated.
We calculated the dipole cross section 
with the gluon effective mass $0.15\,GeV$ (to incorporate confinement)
and $\alpha_s=0.4$ and found $C=2.3$ at $\rho=1/b_0$.
Thus, the energy dependent term in (\ref{6.8}) is fully determined.

The cross section (\ref{6.8}) apparently violates the
Froissart bound and one should perform unitarity 
corrections. Indeed, the partial elastic amplitude 
shows a precocious 
onset of unitarity restrictions at small impact parameters
important even at medium high energies \cite{kpp}.

Following \cite{dl,hp} we assume 
that the $t$-dependence of the $pp$ elastic amplitude 
is given by the Dirac electromagnetic formfactor squared.
Correspondingly, the
mean square radius $\la \tilde r_{ch}^2\ra$ 
evaluated in \cite{hp} 
should be smaller than $\la r_{ch}^2\ra$.

For the dipole parameterization 
of the formfactor the partial elastic amplitude 
which is related via unitarity to
$\sigma^{pp}_n$, given by (\ref{6.4}), (\ref{6.7a}),
takes the form,
\beq
{\rm Im}\,\gamma^{pp}_n(b,s)=
\frac{\sigma^{pp}_n(s)}{8\,\pi\,B_n}\,
y^3\,K_3(y)\ ,
\label{6.14}
\eeq
where $K_3(y)$ is the third order modified Bessel function
and $y=b\sqrt{8/B_n}$. The slope parameter grows
linearly with $n$ due to the random walk of radiated gluons with a
step $1/b_0^2$ in the impact parameter plane,
$B_n = 2\la\tilde r_{ch}^2\ra/3 + n/2b_0^2$.

We unitarize the partial amplitude 
${\rm Im}\,\gamma_P(s,b)=
\sum\limits_{n=0}{\rm Im}\,\gamma_n(s,b)$
using the quasi-eikonal model \cite{kaidalov},
\beq
{\rm Im}\,\Gamma_P(b,s)=
\frac{1 - {\rm exp}
\left[-D(s)\,
{\rm Im}\,\gamma_P(b,s)
\right]}{D(s)}
\label{6.20}
\eeq
where 
$D(s)-1=\sigma_{sd}(s)/\sigma_{el}(s)$
is the ratio of the single diffractive to elastic
cross sections. It is 
approximately equal to $0.25$ at the lowest ISR energy
and slightly decreases with energy $\propto s^{-0.04}$
\cite{schlein,dino}. Note that good results can be also
achieved with a different unitarization scheme similar to 
one suggested in \cite{knp}. The details will be presented 
elsewhere.

In order to calculate the total cross section,
$\sigma_{tot}= 2\int d^2b\,{\rm Im}\Gamma(b,s)$,
one needs to fix 
the energy independent term with $n=0$ in (\ref{6.14}).
This can be 
done comparing with the data for $\sigma_{tot}$ 
at any energy sufficiently
high to neglect Reggeon contributions. We used
the most precise data \cite{cdf} at $\sqrt{s}= 546\,GeV$
and fixed $\tilde\sigma_0=39.7\,mb$.

\begin{figure}[thb]
\includegraphics{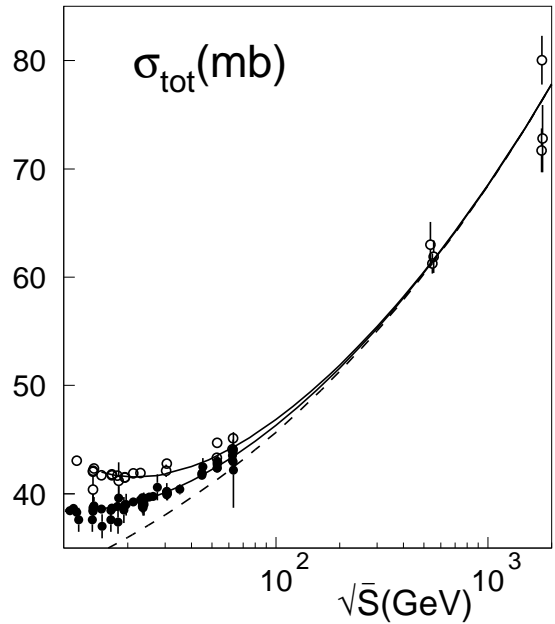}
\begin{center}
\vspace{8.1cm}
\parbox{13cm}
{\caption[shad1]
{Data for total $pp$ (full circles) and $\bar pp$ (open circles)
cross section \cite{pdt}
and the prediction of Eq.~(\ref{6.14}) 
for the energy dependence of
the Pomeron part (dashed curve).
The solid curves 
include Reggeon contributions fitted to the data.}
\label{stot}}
\end{center}
\end{figure}

The predicted energy dependence of $\sigma_{tot}^{pp}$ is shown
by the dashed curve in Fig.~\ref{stot} which is in good
agreement with the data at high energy \cite{pdt}, 
but apparently needs Reggeon corrections towards low energies.
We added a Reggeon term ${\rm Im}\,\Gamma_R(s,b)
[1\,-\,{\rm Im}\,\Gamma_P(s,b)]$
screened by unitarity corrections, which
was fitted independently for $pp$ and $\bar pp$,
$\sigma^{pp}_R=17.8\,mb/\sqrt{s/s_0}$,
$\sigma^{\bar pp}_R=32.8\,mb/\sqrt{s/s_0}$.
The fitted Reggeon slope is $B_R=R_R^2+2\,\alpha_R^{\prime}\,{\rm ln}(s/s_0)$,
where $\alpha_R^{\prime}=0.9\,GeV^{-2}$ and $R^2_R=3\,GeV^{-2}$.

The results are 
shown by the solid curves of Fig.~\ref{stot} ($pp$
bottom curve
and $\bar pp$ upper curve). 

\begin{figure}[tb]
\includegraphics{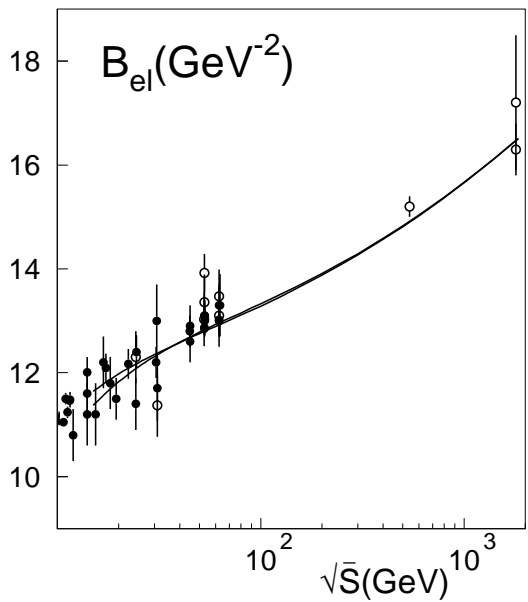}
\begin{center}
\vspace{7.9cm}
\parbox{13cm}
{\caption[shad1]
{Data for the elastic slope \cite{pdt} 
and our predictions. 
The upper and bottom curves and open and full circles 
correspond to $\bar pp$ and $pp$ respectively.}
\label{slope}}
\end{center}
\end{figure}

As soon as the partial amplitude (\ref{6.20}) 
is known, we are in position to predict the
slope of elastic scattering at $t=0$, 
$B_{el}(s) = \la b^2\ra/2$, where averaging is weighted by
the partial amplitude (\ref{6.20}).
The results exhibit good agreement when compared with the $pp$ and $\bar
pp$ data \cite{pdt} in Fig.~\ref{slope}. 
Although the value of the slope essentially depends on
our choice of $\la\tilde r^2_{ch}\ra$ in (\ref{6.14}),
the predicted energy dependence, {\it i.e.}
the effective value $\alpha_P^{\prime}$ is fully defined
by the parameter $b_0$ fixed in \cite{kst2}.
Indeed, each radiated gluon makes a 
``step'' $\sim 1/b_0^2=(0.3\,fm)^2$
in the impact parameter plane leading to the rising
energy dependence of the elastic slope.
Eventually, at very high energies the approximation
of small gluon clouds breaks down but the mean 
number of gluons in a quark 
$\la n\ra = \Delta\,{\rm ln}(s/s_0)$ remains quite small
even in the energy range of colliders. 
It is only $\la n\ra = 0.7-1$ at the ISR 
and reaches about two gluons at the Tevatron.
Correspondingly, the mean square of the quark
radius grows from $0.06\,fm^2$ to  $0.18\,fm^2$ which
is still rather small compared to the mean square of 
the charge radius of the proton.

Summarizing, the strong nonperturbative interaction of radiated gluons
substantially shrinks the gluon clouds around of valence
quarks. This spots are small ($\sim 0.3\,fm$) compared to 
the hadronic radius, but the gluon radiation
grows with energy as $s^{\Delta}$
where $\Delta=0.17\pm 0.01$. Such a steep rise does not
contradict the data since this fraction of the total cross section
is rather small (it contains a factor $1/b_0^2\approx 1\,mb$).
A large energy independent fraction comes from the Born term
which corresponds to scattering of the valence quark skeleton 
without gluon radiation. A very soft interaction which cannot
resolve and excite the small spots contributes to this term.
It cannot
be reliably predicted and is fixed by data , while the energy 
dependent term is fully calculated. The results are in 
good agreement with the data for total $pp$ and $\bar pp$
cross sections and elastic
slopes.

Note that although we have some room for fine-tuning in the 
the parameters ($C,\ s_0,\ \la\tilde r^2_{ch}\ra$),
the results are rather insensitive and the agreement with
data is always pretty good. We have also tried a different unitarization 
scheme suggested in \cite{knp} arriving to similar results.

The details of calculations and further comparison with 
elastic scattering data will be published elsewhere.

{\bf Acknowledgments:} We are thankful to 
J\"org H\"ufner, Andreas Sch\"afer and Sasha Tarasov for 
illuminating and very helpful discussions. We are grateful to
J\"org Raufeisen who has read the paper and made many improving 
comments.
This work was partially supported by the grant 
INTAS-97-OPEN-31696, by the European Network:
Hadronic Physics with Electromagnetic Probes,
Contract No. FMRX-CT96-0008 and by the INFN and MURST of Italy.

\end{document}